\documentclass[twocolumn,showpacs,preprintnumbers,floatfix,amsmath,amssymb,superscriptadd
ress]{revtex4}

\usepackage{graphicx}
\usepackage{dcolumn}
\usepackage{bm}

\newcommand{\be}{\begin{equation}}
\newcommand{\ee}{\end{equation}}
\newcommand{\bn}{\begin{eqnarray}}
\newcommand{\en}{\end{eqnarray}}
\newcommand{\ba}{\begin{array}}
\newcommand{\ea}{\end{array}}
\newcommand{\bc}{\begin{center}}
\newcommand{\ec}{\end{center}}
\newcommand{\bml}{\begin{mathletters}}
\newcommand{\eml}{\end{mathletters}}

\begin{document}

\preprint{ }

\title{High-spin intruder states in the $fp$ shell nuclei and 
isoscalar proton-neutron correlations
}

\author{G. Stoitcheva}

\affiliation{
Physics Division, Oak Ridge National Laboratory, Oak Ridge, Tennessee 37831
}%
\affiliation{
Department of Physics and Astronomy, University of Tennessee, Knoxville, Tennessee
37996
}%

\author{W. Satu{\l}a}
\affiliation{
Institute of Theoretical Physics, University of Warsaw, ul. Ho\.za 69,
00-681 Warsaw, Poland }%
\affiliation{Joint Institute for Heavy Ion Research,
Oak Ridge National Laboratory, P.O. Box 2008, Oak Ridge, Tennessee 37831}

\author{W. Nazarewicz}
\affiliation{
Department of Physics and Astronomy, University of Tennessee, Knoxville, Tennessee
37996
}%
\affiliation{
Physics Division, Oak Ridge National Laboratory, Oak Ridge, Tennessee 37831
}%
\affiliation{
Institute of Theoretical Physics, University of Warsaw, ul. Ho\.za 69,
00-681 Warsaw, Poland }%

\author{D.J. Dean}%
\affiliation{
Physics Division, Oak Ridge National Laboratory, Oak Ridge, Tennessee 37831
}%

\author{M. Zalewski}
\affiliation{
Institute of Theoretical Physics, University of Warsaw, ul. Ho\.za 69,
00-681 Warsaw, Poland }%

\author{H. Zdu\'nczuk}
\affiliation{
Institute of Theoretical Physics, University of Warsaw, ul. Ho\.za 69,
00-681 Warsaw, Poland }%

\date{\today}

\begin{abstract}
We perform a systematic shell-model and mean-field study of 
fully-aligned,  high-spin 
$f_{7/2}^n$ seniority isomers
and $d_{3/2}^{-1} f_{7/2}^{n+1}$ intruder
states in the $A$$\sim$44 nuclei
from the lower-$fp$ shell. The shell-model
calculations are performed in the 
full  $sdfp$ configuration space allowing 
1p-1h
cross-shell excitations. The self-consistent mean-field
calculations are based on the Hartree-Fock approach with the Skyrme
energy density functional that reproduces empirical Landau parameters.
While  there is a  nice agreement between 
experimental and theoretical relative energies of fully-aligned
states in $N$$>$$Z$ nuclei, this is no longer the case for
the $N$=$Z$ systems. The remaining deviation from the data is
attributed to the isoscalar proton-neutron correlations. It is also
demonstrated that the Coulomb corrections at high spins noticeably
depend on the choice of the energy density functional.
\end{abstract}


\maketitle

The triumph  of the nuclear shell model (SM)
in the $sdfp$-shell region \cite{[Pov01],[Cau04],[Hon04],[Dea04]}
proves that many
 spectroscopic properties of those fairly heavy  nuclei  can be  well
accounted for by an  effective two-body $G$-matrix 
augmented by the  monopole corrections. 
In spite of the success of the SM description, 
there are still many open questions and challenges in 
this region  of the nuclear chart that 
offer many opportunities for new physics.
In particular, 
studies  of mirror-symmetric
nuclei and precise measurements of the Coulomb energy displacement
shed light on  isospin breaking effects \cite{[Zuk02],[Bra02]}. 
Another frontier is  investigations of 
unnatural-parity 
intruder states in  $A$$\sim$44 nuclei
from the lower-$fp$ shell
associated with cross-shell excitations across the
$N$=$Z$=20 magic gap that give rise to
shape coexistence effects and emergence of collective rotational
excitations \cite{[Rop04],[Bed04],[Lac05]}. 

From a theoretical standpoint,
the  $fp$-shell nuclei  are particularly
good candidates to study  the competition between 
collective and single-particle excitations.
Since the associated configuration spaces 
are  not prohibitively large for  SM calculations,
and, at the same time, the number of valence particles (and holes)
is  large enough to create substantial collectivity,
these systems 
form a crucial  playground to
confront the  spherical SM  with
collective approaches based on  the mean-field (MF) theory
\cite{[Cau95],[Bed97]}.
The diversity of nuclear structure phenomena,  rich
amount of spectroscopic data collected,
and  a possibility of direct SM verification of MF calculations
in the $fp$  region, offer a unique
opportunity for fine tuning of the underlying nuclear energy
density functional (EDF).

Recently,  a systematic MF
analysis of maximum-spin states (also
referred to as  terminating
states or seniority isomers) has been  performed
within the Skyrme-Hartree-Fock (SHF) approach
\cite{[Zdu05x],[Sat05b]}
for the 
$[f_{7/2}^n]_{I_{max}}$ and $[d_{3/2}^{-1}
f_{7/2}^{n+1}]_{I_{max}}$ configurations ($n$ denotes the number of
valence particles outside the $^{40}$Ca core). 
Those fully-aligned states,
experimentally known   in a number of $20\leq
Z < N\leq 24$ nuclei,
 have fairly simple SM configurations,
and they provide an excellent  testing ground for the time-odd 
densities and fields that appear in the MF description. 
In this context, 
the energy
difference between the excitation energies of the terminating states,
\begin{equation}\label{deltae}
\Delta E = E([d_{3/2}^{-1} f_{7/2}^{n+1}]_{I_{max}}) -
E([f_{7/2}^n]_{I_{max}}),
\end{equation}
  is a sensitive probe of
time-odd spin couplings and the strength of the spin-orbit term in the EDF. 
In particular, it was
demonstrated \cite{[Zdu05x],[Sat05b]} that by constraining the 
Skyrme EDF
to the empirical spin-isospin Landau  parameters
and by slightly  reducing  the  spin-orbit strength, 
good agreement with  the data could  be obtained. 
This result, based on high-spin data for terminating states, 
is consistent with conclusions of previous works
\cite{[Ost92],[Ben02a]} based on different theoretical methodology and 
experimental input 
(such as giant resonances, beta decays, and moments of inertia).

%
\begin{figure}[htb]
    \centering
   \includegraphics[width=8.3cm]{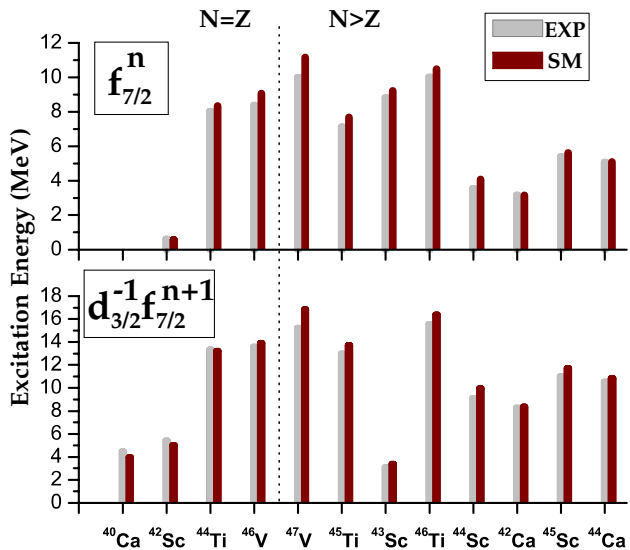}
\hspace{-0.3cm}
\caption{Experimental  (gray bars) and  shell model 
(black bars)  excitation energies of maximum-spin states of
$f_{7/2}^n$ (top) and  $d_{3/2}^{-1}f_{7/2}^{n+1}$ (bottom) configurations
in $N=Z$ (left) and $N >Z$ (right) $fp$-shell nuclei.
Experimental data are taken from Refs.~\cite{[Zdu05x],[Chi05]}.
}\label{shell-m}
\end{figure}
%

The MF studies of
Refs.~\cite{[Zdu05x],[Sat05]} rely on the assumption that the
terminating states are almost ideal examples of unperturbed
single-particle motion, which further implies that $\Delta E$, unlike the
absolute excitation energies $E([d_{3/2}^{-1}f_{7/2}^{n+1}]_{I_{max}})$ and
$E([f_{7/2}^n]_{I_{max}})$, mainly depends 
on properties of the underlying  MF. In particular, $\Delta E$ depends  on
the energy of the cross-shell excitation and symmetry-breaking effects.
The main  objective of the present work is to (i)
study the role of dynamical correlations  on $\Delta E$,
and (ii)  investigate the origin of  large deviations between 
MF results and experimental data  for $N$=$Z$ nuclei. For this purpose,
we carry  SM calculations. 
Preliminary results of this
analysis were published in Refs.~\cite{[Sto05a],[Sat05d]}.

Our SM calculations  were carried out  using the
code ANTOINE~\cite{[Cau95]} in the $sdfp$ configuration
space limited to 1p-1h cross-shell excitation
from the $sd$ shell to the $fp$ shell.
In  the $fp$-shell SM space 
we took  the FPD6 interaction \cite{[Ric90]}. The remaining matrix
elements are those of Ref.~\cite{[War90]}.
As compared to the earlier work~\cite{[Bed97]},
the mass scaling of the SM matrix elements was done here
consistently, thus reducing  the $sd$ interaction channel by $\sim$4\%.
As seen in Fig.~\ref{shell-m},  excellent agreement was obtained
between the SM and experiment for the absolute excitation energies of
terminating states for both $E([f_{7/2}^n]_{I_{max}})$ and
$E([d_{3/2}^{-1}f_{7/2}^{n+1}]_{I_{max}})$ configurations.

%
\begin{figure}[htb]
\begin{center}
\includegraphics[width=8.2cm]{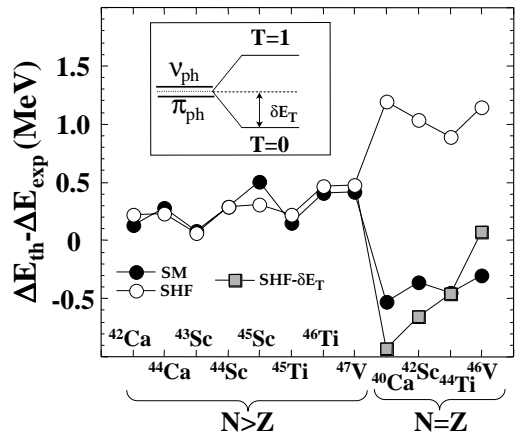}
\caption[]{Difference  $\Delta E_{\rm th} - \Delta E_{\rm exp}$
between experimental and theoretical values of 
$\Delta E$ (\ref{deltae})
in A$\sim$44 mass region.
Dots denote the SM results.
Circles denote the SHF results based on the
modified  SkO
parameterization (see text).
The SHF calculations for the  
 $[d_{3/2}^{-1}f_{7/2}^{n+1}]_{I_{max}}$ intruders in
$N$=$Z$ nuclei yield two nearly degenerate states associated with
proton ($\pi_{\rm ph}$) and neutron ($\nu_{\rm ph}$) cross-shell
excitations. As shown in the inset,
the physical $T$=0 state in the laboratory frame
is shifted down  in energy by  $\delta E_T$ (isospin
correlation energy).
Squares denote the SHF results for $N$=$Z$ nuclei
with the isospin correction added.
The SHF results were shifted by 480\,keV in order
to facilitate the comparison with SM. 
}\label{sko}
\end{center}
\end{figure}

The calculated SM and SHF energy differences $\Delta E$
are  shown in
Fig.~\ref{sko}
relative to experimental values. 
We  note that while Fig.~\ref{shell-m} suggests a
similar level of agreement
between experiment and the SM in $N$=$Z$ and $N$$>$$Z$ nuclei,
the energy differences tell a  different story. Indeed,
in $N$$>$$Z$ nuclei  the SM systematically overestimates the experimental
data by $\sim$280\,keV. On the contrary,
in $N$=$Z$ nuclei the SM systematically underestimates the data
by $\sim$410\,keV. This 
clearly suggests that important correlations related to
isospin and cross-shell excitations  are missing in the
present SM implementation.

The SM results are further compared to the
SHF calculations based on the  
SkO~\cite{[Rei99]} parameterization slightly modified
along the prescription given in Refs.~\cite{[Zdu05x],[Sat05]}.
Without entering into details,
we  recall that the modifications concern coupling constants related
to the time-odd spin fields $C^s_t {\boldsymbol s}^2$
and $C^{\Delta s}_t {\boldsymbol s}$$\cdot$$\Delta{\boldsymbol s}$
where $t=0,1$ labels isoscalar and isovector terms, respectively. 
Moreover, the strength of the
spin-orbit interaction was reduced by 5\%  compared to the original 
SkO value.

\begin{figure}[htb]
\begin{center}
\includegraphics[width=8.2cm]{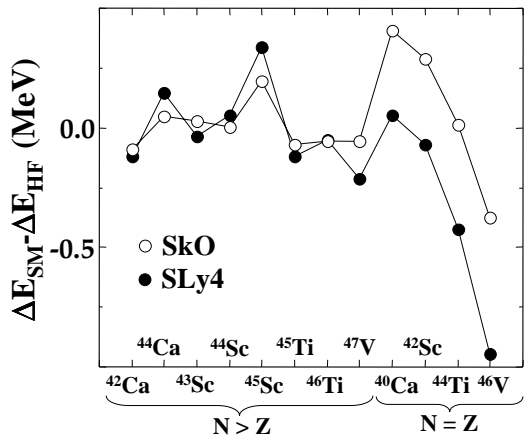}
\caption[]{Difference between SM and SHF values of $\Delta E$.
Two Skyrme parametrizations are used:  SkO (dots) and
SLy4 (circles), modified according to Ref.~\cite{[Zdu05x]}. 
As in Fig.~\ref{sko}, the SHF results
were shifted by 480\,keV. }\label{sl4}
\end{center}
\end{figure}

In contrast to the SM, the SHF underestimates experimental
values of $\Delta E$ in $N$$>$$Z$ nuclei by  $\sim$200\,keV giving rise
to an average offset of $\sim$480\,keV between the two models.
In order to facilitate the comparison, this average difference
was removed by shifting up the HF results in Fig.~\ref{sko}.
(It is to be noted that an overall shift in $\Delta E$ 
 can easily be accounted for by varying the size of 
the $N$=$Z$=20 gap 
in  SM or by a changing the magnitude of the spin-orbit term in SHF.)
It is striking to see that
SHF calculations follow  SM results in $N$$>$$Z$ nuclei extremely well,
reproducing  details of isotopic and isotonic dependence.
This result appears to be fairly general. Indeed,
as seen in Fig.~\ref{sl4},
similar agreement  was obtained
for SHF calculations based on the 
SLy4 parameterization \cite{[Cha97]},
modified according to Ref.~\cite{[Zdu05x]}. 
These results strongly support our assumption
that the maximally-aligned  states in $N$$>$$Z$ nuclei are
excellent examples of an almost unperturbed single-particle
motion and that dynamical correlations present in these states 
do not exhibit any  distinct particle number  dependence.

%

The  difference in the SHF  description of $N$$>$$Z$ and
$N$=$Z$ nuclei seen in Fig.~\ref{sko} can be partly explained in terms
of the  spontaneous breaking of isobaric symmetry
in the $d_{3/2}^{-1} f_{7/2}^{n+1}$ terminating states
in $N$=$Z$ nuclei.  In the MF picture, 
those states are not uniquely defined. Indeed,
by making either neutron ($\nu$) or proton ($\pi$)
$d_{3/2}$$\rightarrow$$f_{7/2}$ 1p-1h excitation, one arrives at
 two nearly degenerate intrinsic states
$E([d_{3/2}^{-1} f_{7/2}^{n+1}]_{I_{max}}^\nu)
\approx E([d_{3/2}^{-1} f_{7/2}^{n+1}]_{I_{max}}^\pi)$, which
manifestly violate isobaric symmetry. Indeed,
these MF states are not eigenstates of isospin. After isospin projection,
  the $T$=0 state  becomes lower in energy
in the laboratory
system,  as illustrated
in the inset of Fig.~\ref{sko}.
Due to physical symmetry-breaking caused
 by the Coulomb interaction, the two 
intrinsic states are
slightly split with the proton 1p-1h excitation being
always slightly lower in energy.
The reason is that proton excitation from the
 $d_{3/2}$ orbit to a more extended $f_{7/2}$ orbit
slightly increases  the mean charge radius, thus reducing the Coulomb
repulsion.

%
\begin{figure}[htb]
\begin{center}
\includegraphics[width=8.2cm]{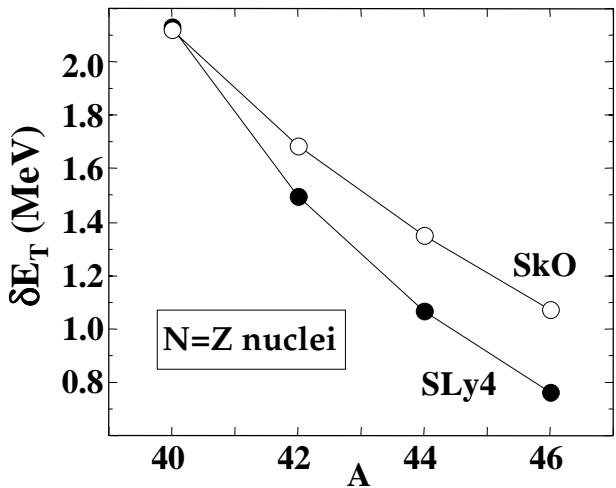}
\caption[]{Phenomenological estimates
of the isospin energy correction, $\delta E_T (A)$, due to the
restoration of isobaric symmetry internally 
broken  in SHF  solutions corresponding
to the $[d_{3/2}^{-1} f_{7/2}^{n+1}]_{I_{max}}$ terminating states
in $N$=$Z$ nuclei. The values of $\delta E_T (A)$ obtained 
in SkO and SLy4 models
are labeled by open and filled dots,
respectively.
}\label{et}
\end{center}
\end{figure}

In order to make comparison to the data,
the correlation energy $\delta E_T$ due to isospin symmetry-breaking
 in SHF should be estimated.
For the purpose of this work, we  evaluate 
$\delta E_T$ using the  self-consistent isocranking \cite{[Sat01x]}.
That is,
we  compute the energy difference between
the isobaric analogue states at high spin, i.e.,
$\delta E_T \equiv E([d_{3/2}^{-1} f_{7/2}^{n+1}]_{I_{max}}; T_z=\pm 1) -
E([d_{3/2}^{-1} f_{7/2}^{n+1}]_{I_{max}}; T_z=0)$
using the SHF approach with  Coulomb interaction switched off.
The energy difference $\Delta E^{(T=0)}_{HF}$ corrected in this way is
marked by squares in Fig.~\ref{sko}.

The calculated isospin
corrections  are depicted
in Fig.~\ref{et}. It is interesting to observe that $\delta E_T (A)$
 shows a surprisingly strong decrease with
increasing $A$. According to our analysis,
this strong particle-number dependence can be attributed
to the time-odd fields, and the
calculations indicate
 that this effect can be reduced by decreasing
the value of the
isovector Landau parameter $g^\prime_0$. Whether or not this can be
used to further constrain the value of
$g^\prime_0$ remains to be studied (see, however,
 recent work \cite{[Sat03],[Sat05c]}).
Coming back to Figs.~\ref{sko} and \ref{sl4}, it is encouraging to see
that after approximate
isospin symmetry restoration, one obtains
 $\Delta E^{(T=0)}_{\rm SHF} \approx \Delta
E^{(T=0)}_{\rm SM}$ also in $N$=$Z$ nuclei.
Hence, our comparative study strongly suggests that
correlations of a similar type  are missing in $N$=$Z$ nuclei,
both in the SM and the  SHF approaches.

\begin{figure}[hbt]
\begin{center}
\includegraphics[width=8.2cm]{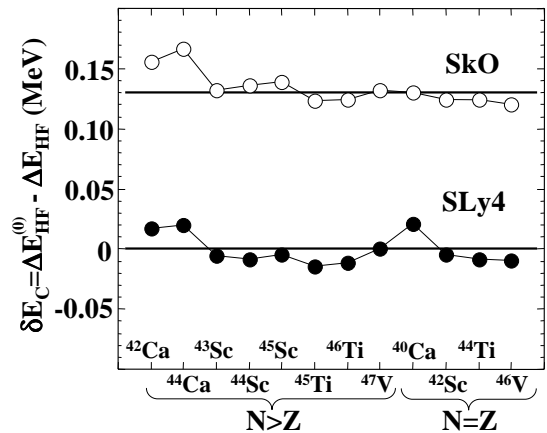}
\caption[]{Coulomb correction,  $\delta E_C$,  to $\Delta E$
calculated in the SkO and SLy4 models by performing SHF calculations without
($\Delta E_{HF}^{(0)}$)  and with ($\Delta E_{HF}$)
 Coulomb interaction.
}\label{coul}
\end{center}
\end{figure}

Our SM interaction conserves isospin. Consequently, 
the Coulomb correction
to $\Delta E$, $\delta
E_C$ should be added afterwards. The Coulomb correction (including
the associated  isovector polarization) can be
calculated self-consistently in  SHF.
Surprisingly, the many-body response  against electrostatic 
polarization 
appears to be strongly sensitive to the isovector part of the EDF.
This is visualized in Fig.~\ref{coul} which shows  a
difference, $\delta E_C$, between the SHF values of $\Delta E$ calculated 
 without ($\Delta E_{HF}^{(0)}$) and with
($\Delta E_{HF}$)  the Coulomb term.
While   $\delta E_C$ is  very small for  SLy4,
the values calculated in the 
SkO variant are appreciable,
$\delta E_C\approx 130$\,keV. 
The difference can be traced back to the fact that
these two parameterizations differ strongly in  the strength of the
isovector part of the spin-orbit interaction. While in SLy4 the ratio of
the isovector ($W_1$) to the isoscalar ($W_0$) spin-orbit strengths
equals to the standard value of $W_1/W_0 = 1/3$, SkO is a modern
parameterization having $W_1/W_0 \approx -1.3$. The resulting change in
the radial form factor leads to a large Coulomb effect at high spin, an
effect that  is of the same order as the measured Coulomb energy
differences in $fp$-shell nuclei \cite{[Bra02]}. 
Based on our study,  the  Coulomb interaction can give rise to an overall 
displacement of the order of 100\,keV that very weakly depends on $Z$
and $N$.

%

In summary, the self-consistent SHF analysis  of terminating states in
the $A$$\sim$44 nuclei agrees nicely with  SM studies, after correcting
the former for the isospin-breaking effects in $N$=$Z$ nuclei. For 
$N$$>$$Z$
nuclei, 
both theories provide a good reproduction of experimental data.
This validates the assumption of previous studies
\cite{[Zdu05x],[Sat05b]} regarding the single-particle character of the
maximally-aligned states. We believe that the origin of the remaining
deviation from the data seen in  the $N$=$Z$ systems has its source in the
$T$=0 pairing channel. The  SM provides an excellent description
of spectroscopic properties in the the whole $fp$ shell. Therefore, any
discrepancy involving intruder configurations must have its source in
the assumed  truncation to 1p-1h cross-shell excitations. This
configuration-space restriction
is expected to  impact the  isoscalar  channel
associated with the $sd$$\rightarrow$$fp$ pair scattering. The single
main obstacle that prevents us from carrying out calculations in an
extended space involving 2p-2h, 3p-3h,..., cross-shell transitions is
the lack of an  appropriate effective interaction. On the SHF level,
while the extended proton-neutron self-consistent SHF formalism has been
developed \cite{[Per04]},  its practical implementation is still in  an
early stage; see, e.g., Refs. \cite{[Per04],[Ter98]} and references
quoted therein.  

Finally, we have demonstrated that the state-dependent
Coulomb polarization  at high spins noticeably depends on the choice of
the energy density functional, in particular its spin-dependent terms.
The resulting uncertainty in the Coulomb energy shift can be as large as
the measured Coulomb energy displacement. This is likely to result in 
ambiguities when estimating   Coulomb effects at high spins.

\begin{acknowledgments} 
This work was supported in part by  the U.S. Department of Energy
under Contracts Nos. DE-FG02-96ER40963 (University of Tennessee),
DE-AC05-00OR22725 with UT-Battelle, LLC (Oak Ridge National
Laboratory), DE-FG05-87ER40361 (Joint Institute for Heavy Ion
Research);  by the Polish Committee for
Scientific Research (KBN) under contract No. 1~P03B~059~27; and  by the
Foundation for Polish Science (FNP).
\end{acknowledgments}


\begin{thebibliography}{10}

\bibitem{[Pov01]}
{A. Poves, J. Sanchez-Solano, E. Caurier, and F. Nowacki, Nucl. Phys. A {\bf
  694}, 157 (2001)}.

\bibitem{[Cau04]}
{E. Caurier, F. Nowacki, and A. Poves, Nucl. Phys. A {\bf 742}, 14 (2004)}.

\bibitem{[Hon04]}
{M. Honma, T. Otsuka, B.A. Brown, and T. Mizusaki, Phys. Rev. C {\bf 69},
  034335 (2004)}.
  
\bibitem{[Dea04]}
{D.J. Dean {\it et al.}, Prog. Part. Nucl. Phys.  {\bf 53}, 419 (2004)}.  

\bibitem{[Zuk02]}
{A.P. Zuker, S.M. Lenzi, G. Martinez-Pinedo, A. Poves, Phys. Rev. Lett. {\bf
  89}, 142502 (2002)}.

\bibitem{[Bra02]}
{F. Brandolini {\it et al.}, Phys. Rev. C {\bf 66}, 021302 (2002)}.

\bibitem{[Rop04]}
{H. R\"opke, Eur. Phys. J. A {\bf 22}, 213 (2004)}.

\bibitem{[Bed04]}
{P. Bednarczyk {\it et al.}, Eur. Phys. J. A {\bf 20}, 45 (2004)}.

\bibitem{[Lac05]}
{M. Lach {\it et al.}, Eur. Phys. J. A {\bf 25}, 1 (2005)}.

\bibitem{[Cau95]}
{E. Caurier {\it et al.}, Phys. Rev. Lett. {\bf 75}, 2466 (1995).}

\bibitem{[Bed97]}
{P. Bednarczyk {\it et al.}, Phys. Lett. B {\bf 393}, 285 (1997).}

\bibitem{[Zdu05x]}
{H. Zdu{\'n}czuk, W. Satu{\l}a, and R. Wyss, Phys. Rev. C {\bf 58} 024305
  (2005); Int. J. Mod. Phys. E {\bf 14}, 451 (2005)}.

\bibitem{[Sat05b]}
{W. Satu{\l}a, R. Wyss, and H. Zdu\'nczuk, Eur. Phys. J. A Direct (2005)
  /epjad/i2005-06-067-3.}

\bibitem{[Ost92]}
{F. Osterfeld, Rev. Mod. Phys. {\bf 64} , 491 (1992).}

\bibitem{[Ben02a]}
{M. Bender {\it et al.\/}, Phys. Rev. C {\bf 65}, 054322 (2002).}

\bibitem{[Chi05]}
{C.J. Chiara, private communication, 2005; to be published}.

\bibitem{[Sat05]}
{W. Satu{\l}a and R. Wyss, Rep. Prog. Phys. {\bf 68}, 131 (2005).}

\bibitem{[Sto05a]}
{G. Stoitcheva,  W. Nazarewicz, and D.J. Dean,  
  Eur. Phys. J. A Direct (2005)
  /epjad/i2005-06-207-9.}

\bibitem{[Sat05d]}
{W. Satu{\l}a, Phys. Scripta (2005) in print; nucl-th/0508066.}

\bibitem{[Ric90]} W.A. Richter {\it et al.\/},
Nucl. Phys. A {\bf 523}, 325 (1991).  

\bibitem{[War90]} E.K. Warburton, J.A. Becker, and B.A. Brown, 
		Phys. Rev. C {\bf 41}, 1147 (1990).


\bibitem{[Rei99]}
{P.-G. Reinhard {\it et al.\/}, Phys. Rev. C {\bf 60}, 014316 (1999).}

\bibitem{[Cha97]}
{E. Chabanat, {\it et al.\/}, Nucl. Phys. A {\bf 627}, 710 (1997); 
Nucl. Phys. A {\bf 635}, 231 (1998).}

\bibitem{[Sat01x]}
{W. Satu{\l}a and R. Wyss, Phys. Rev. Lett. {\bf 86}, 4488 (2001); 
{\bf 87} 052504 (2001).}

\bibitem{[Sat03]}
{W. Satu{\l}a and R. Wyss, Phys. Lett. B {\bf 572}, 152 (2003).}

\bibitem{[Sat05c]}
{W. Satu{\l}a, R. Wyss, and M. Rafalski, submitted to Phys. Rev. Lett. 
(2005);   nucl-th/0508004.}

\bibitem{[Per04]}
{E. Perli\'nska, S.G. Rohozi\'nski, J. Dobaczewski, and W. Nazarewicz, 
Phys.   Rev. C {\bf 69}, 014316 (2004)}.

\bibitem{[Ter98]}
{ J. Terasaki, R. Wyss, and P.-H. Heenen, Phys. Lett. B {\bf 437}, 1 (1998).}

\end{thebibliography}

\end{document}